# THE ENERGY-MOMENTUM TENSOR, THE TRACE IDENTITY AND THE CASIMIR EFFECT

S.G.Kamath *

Department of Mathematics, Indian Institute of Technology Madras,
Chennai 600 036,India

Abstract:

The trace identity associated with the scale transformation $x^\mu \to x'^\mu = e^{-\rho} x^\mu$ on the Lagrangian density for the noninteracting electromagnetic field in the covariant gauge is shown to be violated on a single plate on which the Dirichlet boundary condition $A^\mu \left(t, x^1, x^2, x^3 = -a\right) = 0$ is imposed. It is however respected in free space, i.e. in the absence of the plate; these results reinforce our assertions in an earlier paper where the same exercise was carried out using the Lagrangian density for the free, massive, real scalar field in 2 + 1 dimensions.

*e-mail: kamath@iitm.ac.in

# 1. Introduction

The trace identity [1] associated with the scale transformation $x^\mu \to x'^\mu = e^{-\rho} x^\mu$ on a scalar field $\phi(x)$, for example, is conventionally expressed in terms of the energy-momentum tensor $\hat{\Theta}^{\mu\nu}$, and for the massless free real scalar field Lagrangian density in 3 + 1 dimensions, namely, $L = \frac{1}{2} \partial^\alpha \phi \, \partial_\alpha \phi$ the trace identity is given by

$$g_{\mu\nu} \Gamma^{\mu\nu}(y; x, z) + id \left( \delta^{(4)}(y-x) + \delta^{(4)}(y-z) \right) G(x, z) = 0 \qquad (1)$$

with the canonical scale dimension d = 1; G ( x , z ) and $\Gamma^{\mu\nu}(y; x, z)$ being the connected T *-ordered products $\langle 0 | T^* \left( \phi(x)\phi(z) \right) | 0 \rangle_c$ and $\langle 0 | T^* \left( \hat{\Theta}^{\mu\nu}(y)\phi(x)\phi(z) \right) | 0 \rangle_c$ respectively. $\hat{\Theta}^{\mu\nu}$ is defined [1] in terms of the canonical energy-momentum tensor $\hat{\Theta}^{\mu\nu}_c = \partial^\mu \phi \, \partial^\nu \phi - g^{\mu\nu} L$ by

$$\hat{\Theta}^{\mu\nu} = \hat{\Theta}^{\mu\nu}_c + \frac{1}{6} \left( g^{\mu\nu} \partial^\alpha \partial_\alpha - \partial^\mu \partial^\nu \right) \phi^2 \qquad (2)$$

with the canonical dilatation current

$$D^\lambda_c = x_\mu \hat{\Theta}^{\lambda\mu}_c + \frac{1}{2} \partial^\lambda \phi^2 \qquad (3)$$

As emphasized by Coleman and Jackiw[1], one can now rework (3) to get a modified version of the dilatation current given by

$$D^\mu = x_\lambda \hat{\Theta}^{\mu\lambda} \qquad (4)$$

as the second term defining the field virial $V^\lambda$ is a total divergence.

For the free electromagnetic field on the other hand, it is well known [2] that the corresponding $\hat{\Theta}^{\mu\nu}$ associated with the Lagrangian density

$$L = -\frac{1}{4} F_{\alpha\beta} F^{\alpha\beta} , \quad F^{\alpha\beta} = \partial^\alpha A^\beta - \partial^\beta A^\alpha \tag{5}$$

is the Belinfante tensor $\hat{\Theta}_B^{\mu\nu}$, the latter being defined [1] in terms of the canonical energy-momentum tensor $\hat{\Theta}_c^{\mu\nu}$ by

$$\hat{\Theta}_B^{\mu\nu} = \hat{\Theta}_c^{\mu\nu} + \partial_\alpha X^{\alpha\mu\nu} \tag{6}$$

with

$$\hat{\Theta}_c^{\mu\nu} = \Pi^{\mu\alpha} \partial_\alpha A^\nu - g^{\mu\nu} L \tag{6a}$$

and
$$X^{\alpha\mu\nu} = \frac{1}{2} \left( \Pi^{\alpha\beta} \Sigma_{\beta\lambda}^{\mu\nu} - \Pi^{\mu\beta} \Sigma_{\beta\lambda}^{\alpha\nu} - \Pi^{\nu\beta} \Sigma_{\beta\lambda}^{\alpha\mu} \right) A^\lambda \tag{6b}$$

the counterpart of (4) now being

$$D^\alpha = x_\beta \hat{\Theta}_B^{\alpha\beta} \tag{7}$$

In eqs.(6) $\Pi^{\alpha\beta} = -F^{\alpha\beta}$ and $\Sigma_{\beta\lambda}^{\mu\nu} = \delta_\beta^\mu \delta_\lambda^\nu - \delta_\beta^\nu \delta_\lambda^\mu$.

This paper extends the derivation leading to eqs.(6) and (7) and associated with the Coulomb gauge Lagrangian (5) to the Lagrangian density for the free electromagnetic field in the covariant gauge given by

$$L = -\frac{1}{4} F_{\alpha\beta} F^{\alpha\beta} - \frac{1}{2\lambda} \left( \partial^\mu A_\mu \right)^2 \tag{8}$$

While this is a routine exercise using the methods of Coleman and Jackiw[1] it has been included here - with the relevant steps gone through in some detail - merely for the sake of completeness; more importantly, for the issues which this paper addresses and which motivate this paper, the relevant energy-momentum tensor is not the Belinfante version $\hat{\Theta}_B^{\mu\nu}$ but the energy - momentum tensor $\hat{\Theta}^{\mu\nu}$ defined by [3]

$$\hat{\Theta}^{\mu\nu} = \hat{\Theta}_B^{\mu\nu} + \frac{1}{2}\partial_\alpha \partial_\beta X^{\alpha\beta\mu\nu} \tag{9a}$$

where

$$X^{\alpha\beta\mu\nu} = g^{\alpha\beta}\sigma^{\mu\nu} - g^{\alpha\mu}\sigma^{\beta\nu} - g^{\alpha\nu}\sigma^{\beta\mu} + g^{\mu\nu}\sigma^{\alpha\beta} - \frac{1}{3}\left(g^{\alpha\beta}g^{\mu\nu} - g^{\alpha\mu}g^{\beta\nu}\right)\sigma_\lambda^\lambda \tag{9b}$$

with

$$\sigma^{\mu\nu} = \frac{1}{2}g^{\mu\nu}A_\alpha A^\alpha - A^\mu A^\nu, \quad \sigma_\lambda^\lambda = g_{\lambda\mu}\sigma^{\mu\lambda} \tag{9c}$$

As is obvious, $\sigma^{\mu\nu}$ is symmetric in $\mu$ and $\nu$ and its nonzero value underlines the preference of $\hat{\Theta}^{\mu\nu}$ over $\hat{\Theta}_B^{\mu\nu}$ for the Lagrangian in the covariant gauge given by (8).

While relegating these details to the text below, let's motivate this paper with the following question:

Given that the trace identity (see eq.(10) below) associated with the scale transformation $x^\mu \to x'^\mu = e^{-\rho}x^\mu$ on the Lagrangian (8) is maintained, is the same trace identity also maintained on a boundary? To elaborate, we shall first use standard methods in the sequel to check that the trace identity given by

$$g_{\alpha\beta}\langle 0|T^*\left(\hat{\Theta}^{\alpha\beta}(z)A^\mu(s)A^\nu(t)\right)|0\rangle_c + i\delta(z-s)\langle 0|T^*\left(A^\mu(s)A^\nu(t)\right)|0\rangle_c$$
$$+ i\delta(z-t)\langle 0|T^*\left(A^\mu(s)A^\nu(t)\right)|0\rangle_c = 0 \tag{10}$$

is preserved in free space. Subsequently, we check if eq.(10) also holds on a single plate at

$x^3 = $ - a on which the Dirichlet boundary condition $A^\mu(t, x^1, x^2, x^3 = -a) = 0$ is introduced; note that $x^3$ denotes the spatial z-component of the four vector $x^\mu$. For the sake of clarity we shall label these twin checks below as being carried out without and with a boundary respectively. While this exercise is inspired by the Casimir effect [4,5] in which the force between a pair of plates is calculated – indeed, this explains the use of the term in the title of the paper – the objective here is clearly different: Namely to determine how a *minimal* change from an unconstrained configuration represented by the absence of a boundary to a constrained one characterized by the presence of a single plate at $x^3 = -a$ and on which the Dirichlet boundary condition holds $A^\mu \left(t, x^1, x^2, x^3 = -a\right) = 0$ affects, if at all, the validity of the trace identity given by eq.(10) above. Needless to say so, within the above framework the Casimir configuration of a pair of plates shall be understood as a *not - so – minimal* change. Since the objective of this paper has been spelt out above, the calculation of the Casimir energy on the single plate will not engage our attention here, more so as the literature [4,5] on this aspect as well as on the experimental support [6,7] for the Casimir effect is vast, with little or no mention to the best of our knowledge, of the issues raised here. Let's reiterate at this stage that it is the *susceptibility*, if any, of the trace identity given by eq.(10) to a minimal change as explained above from the unconstrained configuration that is on test in this paper. We must also emphasize here that the *departure* from the use of a pair of parallel conducting plates separated by a distance that was used in the original Casimir calculation [8] to a single plate in this paper has been done here deliberately.

The impetus for this exercise comes from an analogous report by the author[9] recently on the validity of the trace identity associated with the Lagrangian density for a noninteracting massive real scalar field viz. $L = \frac{1}{2}\partial^\alpha \phi \partial_\alpha \phi - \frac{1}{2}m^2\phi^2$ in 2 + 1 dimensions when one moves from an unconstrained configuration, wherein the identity has been found to hold, to a constrained configuration – in which the trace identity is shown to be violated - where a boundary in the form of a single plate at $x^2$ = - a is introduced and on which the Dirichlet boundary condition $\phi\left(t, x^1, x^2 = -a\right) = 0$ is imposed. The present work is thus an extension to the Lagrangian given by (8) of the work done in Ref.9.

The plan of this paper is as follows: In Sec. 2 we shall present very briefly following the methods of Ref. 1 a rationale for the choice of the energy-momentum tensor given by eq.(9a) so as to obtain a modified dilatation current in a form analogous to eq.(4) for the Lagrangian for the massless free scalar field and follow it up with an explicit verification of the trace identity given by (10) for the unconstrained configuration; Sec.3 is the main part of this paper wherein we use a modified generating functional following Bordag, Robaschik and Wieczorek[10] which explicitly incorporates the Dirichlet boundary condition $A^\mu\left(t, x^1, x^2, x^3 = -a\right) = 0$ to derive the connected Green's functions relevant for the verification of the trace identity. The paper concludes with a short Appendix that includes a derivation of some of the relevant equations discussed in Sec 3.

## 2. A check of eq.(10) without a boundary

The canonical dilatation current associated with the Lagrangian density (8) is given by

$$D_c^\mu = x_\nu \hat{\Theta}_c^{\mu\nu} + \Pi^{\mu\nu} A_\nu \tag{11a}$$

with $\Pi^{\alpha\beta} = -F^{\alpha\beta} - \frac{1}{\lambda} g^{\alpha\beta}(\partial \cdot A)$. This extra second term in the definition of $\Pi^{\alpha\beta}$ relative to just $\Pi^{\alpha\beta} = -F^{\alpha\beta}$ for the Coulomb gauge Lagrangian (5) now implies that with eqs.(6),(6a) and

$$X^{\alpha\mu\nu} = \left(\Pi^{\alpha\beta} \Sigma^{\mu\nu}_{\beta\lambda} - \Pi^{\mu\beta} \Sigma^{\alpha\nu}_{\beta\lambda} - \Pi^{\nu\beta} \Sigma^{\alpha\mu}_{\beta\lambda}\right) A^\lambda \tag{11b}$$

one can rework eq.(11a) above as

$$D_c^\mu = x_\nu \hat{\Theta}_B^{\mu\nu} + \Pi^{\mu\nu} A_\nu + g_{\alpha\beta} X^{\beta\mu\alpha} - \partial_\beta\left(x_\alpha X^{\beta\mu\alpha}\right) \tag{11c}$$

Note that the r.h.s. of eq.(11b) is now twice that of the r.h.s. of (6b); this difference will turn out to be crucial below. From the second and third terms in (11c) one obtains

$$\begin{aligned}
\Pi^{\mu\nu} A_\nu + g_{\alpha\beta} X^{\beta\mu\alpha} &= F^{\mu\nu} A_\nu + \frac{5}{\lambda}(\partial \cdot A) A^\mu \\
&= \left(\partial^\mu A^\nu - \partial^\nu A^\mu\right) A_\nu + \frac{5}{\lambda}(\partial \cdot A) A^\mu \\
&= \partial_\alpha \sigma^{\mu\alpha}
\end{aligned} \tag{12}$$

when λ = - 5; the last equality defines the field virial $V^\mu = \partial_\alpha \sigma^{\mu\alpha}$, with $\sigma^{\mu\alpha}$ given by (9c) for the covariant gauge Lagrangian density given by (8).We can now use eqs.(9a) and (9b) above to define the dilatation current in terms of $\hat{\Theta}^{\mu\nu}$ as $D^\mu = x_\alpha \hat{\Theta}^{\mu\alpha}$ as in eq.(A.21) of Ref. 1;and then derive the trace identity given in (10) following the steps given in Appendix B of Ref.1.By contrast, retaining the same form as (6b) yields for the l.h.s. of (12)

$$\Pi^{\mu\nu} A_\nu + g_{\alpha\beta} X^{\beta\mu\alpha} = \left( \Pi^{\mu\nu} + \Pi^{\nu\mu} \right) A_\nu + \frac{4}{\lambda}(\partial \cdot A) A^\mu$$

$$= -\frac{2}{\lambda}(\partial \cdot A) A_\nu + \frac{4}{\lambda}(\partial \cdot A) A^\mu = \frac{2}{\lambda}(\partial \cdot A) A_\nu \qquad (13)$$

Unlike the r.h.s. of eq.(12), the r.h.s. of the last step of (13) cannot be written as a total derivative; while this underlines the advantage here in the choice of (11b) over (6b) it also emphasizes the utility of the $F^{\mu\nu} A_\nu$ term in obtaining the desired form of the field virial in (12).Let's also point out here that we do not claim originality in deriving eq.(12) above for the associated value of λ, as we are confident that the steps that have been worked through above must have also been obtained in the unpublished literature.

For the sake of completeness let's now hint at a short derivation of the trace identity given by eq.(10). It is convenient to start from the following Ward identities, firstly for the canonical energy momentum tensor $\hat{\Theta}_c^{\mu\nu}$:

$$\partial_\alpha^{(z)} \langle 0 | T^* \left( \hat{\Theta}_c^{\alpha\beta}(z) A^\mu(s) A^\nu(t) \right) | 0 \rangle_c + i\delta^{(4)}(z-s) \langle 0 | T^* \left( \partial^\beta A^\mu(s) A^\nu(t) \right) | 0 \rangle_c +$$
$$i\delta^{(4)}(z-t) \langle 0 | T^* \left( A^\mu(s) \partial^\beta A^\nu(t) \right) | 0 \rangle_c = 0$$

and secondly for the canonical dilatation current $D_c^\mu$:

$$\partial_\alpha^{(z)} \langle 0 | T^* \left( D_c^\alpha(z) A^\mu(s) A^\nu(t) \right) | 0 \rangle_c + i\delta^{(4)}(z-s) \langle 0 | T^* \left( \delta A^\mu(s) A^\nu(t) \right) | 0 \rangle_c +$$
$$i\delta^{(4)}(z-t) \langle 0 | T^* \left( A^\mu(s) \delta A^\nu(t) \right) | 0 \rangle_c = 0$$

with $\delta A^\mu(x) = (1 + x \cdot \partial) A^\mu(x)$. With the derivation of the dilatation current as $D^\mu = x_\alpha \hat{\Theta}^{\mu\alpha}$ explained in detail in the earlier part of this section it is now a simple matter to repeat verbatim the steps of Appendix B of Ref.1 and obtain the trace identity given by eq.(10) in this paper. As a matter of caution it is useful in this exercise to bear in mind the remarks on p.589 and 590 of Ref.1.

We shall now check the trace identity in the absence of a boundary and begin with the familiar expression for the generating functional for the connected Green's functions that is appropriate for the discussion below, viz.

$$\exp iZ\left[J^\mu, K^{\mu\nu}, M^{\alpha\beta}\right] = N\int DA^\mu \exp i\int_x \left(L + J^\mu A_\mu + K^{\mu\nu}\partial_\mu A_\nu + M^{\alpha\beta}\hat{\Theta}_{\alpha\beta}\right)$$

$$= N\int DA^\mu \left(\sum_0^\infty \frac{i^n}{n!}(M^{\alpha\beta}\hat{\Theta}_{\alpha\beta})^n\right)\exp i\int_x \left(L + J^\mu A_\mu + K^{\mu\nu}\partial_\mu A_\nu\right) \quad (14)$$

where $J^\mu$, $K^{\mu\nu}$ and $M^{\mu\nu}$ are external sources for $A^\mu(x)$, $\partial_\mu A_\nu$ and $\hat{\Theta}_{\mu\nu}$, and N a normalizing constant which ensures that the l.h.s. is unity when the external sources are zero. On expressing $\hat{\Theta}_{\mu\nu}$ in terms of functional derivatives – given further below - acting on the exponential one writes (14) as

$$\exp iZ\left[J^\mu, K^{\mu\nu}, M^{\alpha\beta}\right] = \left(\sum_0^\infty \frac{i^n}{n!}(M^{\alpha\beta}\hat{\Theta}_{\alpha\beta})^n\right) N\int DA^\mu \exp i\int_x \left(L + J^\mu A_\mu + K^{\mu\nu}\partial_\mu A_\nu\right)$$

$$= \left(\sum_0^\infty \frac{i^n}{n!}(M^{\alpha\beta}\hat{\Theta}_{\alpha\beta})^n\right)\exp\left(-\frac{i}{2}\int\int_{x\,y}\left(J_\lambda(x) + iK_{\rho\lambda}\vec{\partial}^\rho_{(x)}\right)D^{\lambda\beta}(x-y)\left(J_\beta(y) + i\vec{\partial}^\alpha_{(y)}K_{\alpha\beta}\right)\right) \quad (15)$$

In eqs.(14) and (15) the symbol $\int_x$ denotes $\int d^4x$, with

$$iD^{\alpha\beta}(x-y) = i\int_k \frac{e^{-ik\cdot(x-y)}}{k^2 + i\varepsilon}\left(6\frac{k^\alpha k^\beta}{k^2} - g^{\alpha\beta}\right) \quad (16)$$

being the propagator associated with the Lagrangian (8) when $\lambda = -5$; also in (16) $\int_k$ is shorthand for $\int \frac{d^4k}{(2\pi)^4}$. Thus

$$i\frac{\delta^2 Z}{\delta J^\alpha(x)\delta J^\beta(y)}\bigg|_{sources=0} = i^2 \langle 0|T^*\left(A^\alpha(x)A^\beta(y)\right)|0\rangle_c = -iD^{\alpha\beta}(x-y) \qquad (17)$$

It is also easy to check that

$$\left(\partial^\mu\partial_\mu g_{\alpha\beta} - \frac{6}{5}\partial_\alpha\partial_\beta\right)D^{\beta\lambda}(x-y) = \delta_\alpha^\lambda \delta^{(4)}(x-y) \qquad (18)$$

the derivative operators being taken with respect to $x^\mu$. We now give below $\hat{\Theta}_{\alpha\beta}$ in terms of functional derivative operators which will act on the exponential on the r.h.s. of eq.(15)

$$\hat{\Theta}_{\alpha\beta}(z) = (-i)^2 \left(C_{\alpha\beta}(z) + S_{\alpha\beta}(z) + T_{\alpha\beta}(z)\right) \qquad (19)$$

with:

$$C_{\alpha\beta} = -\left\{\begin{array}{l} g^{\mu\lambda}\left(\dfrac{\delta}{\delta K^{\alpha\lambda}} - \dfrac{\delta}{\delta K^{\lambda\alpha}} - \dfrac{1}{5}g_{\alpha\lambda}g^{\beta\rho}\dfrac{\delta}{\delta K^{\beta\rho}}\right)\dfrac{\delta}{\delta K^{\beta\lambda}} \\ +g_{\alpha\beta}\left(-\dfrac{1}{4}\left(\dfrac{\delta}{\delta K^{\rho\lambda}} - \dfrac{\delta}{\delta K^{\lambda\rho}}\right)\left(\dfrac{\delta}{\delta K_{\rho\lambda}} - \dfrac{\delta}{\delta K_{\rho\lambda}}\right)\right) + \dfrac{1}{10}\left(g^{\mu\lambda}\dfrac{\delta}{\delta K^{\mu\lambda}}\right)^2 \end{array}\right\}$$

$$S_{\alpha\beta} = \partial^\lambda \left[\left\{\begin{array}{l} -\left(\dfrac{\delta}{\delta K^{\lambda\rho}} - \dfrac{\delta}{\delta K^{\rho\lambda}} - \dfrac{1}{5}g_{\lambda\rho}g^{\tau\nu}\dfrac{\delta}{\delta K^{\tau\nu}}\right)\Sigma^{\rho\mu}_{\alpha\beta} + \left(\dfrac{\delta}{\delta K^{\alpha\rho}} - \dfrac{\delta}{\delta K^{\rho\alpha}} - \dfrac{1}{5}g_{\alpha\rho}g^{\tau\nu}\dfrac{\delta}{\delta K^{\tau\nu}}\right)\Sigma^{\rho\mu}_{\lambda\beta} \\ +\left(\dfrac{\delta}{\delta K^{\beta\rho}} - \dfrac{\delta}{\delta K^{\rho\beta}} - \dfrac{1}{5}g_{\beta\rho}g^{\tau\nu}\dfrac{\delta}{\delta K^{\tau\nu}}\right)\Sigma^{\rho\mu}_{\alpha\lambda} \end{array}\right\}\dfrac{\delta}{\delta J^\mu}\right]$$

$$T_{\alpha\beta} = \frac{1}{2}\partial^\mu \partial^\nu \left\{ \begin{array}{l} g_{\mu\nu}\left(\frac{2}{3}g_{\alpha\beta}\frac{\delta^2}{\delta J^\lambda \delta J_\lambda} - \frac{\delta^2}{\delta J^\alpha \delta J^\beta}\right) - g_{\mu\alpha}\left(\frac{2}{3}g_{\nu\beta}\frac{\delta^2}{\delta J^\lambda \delta J_\lambda} - \frac{\delta^2}{\delta J^\nu \delta J^\beta}\right) \\ + g_{\nu\beta}\frac{\delta^2}{\delta J^\mu \delta J^\alpha} - g_{\alpha\beta}\frac{\delta^2}{\delta J^\mu \delta J^\nu} \end{array} \right\}$$

Note that the partial derivative and functional derivative operators above are taken with respect to $z^\mu$ and should be understood as $\partial^\mu_{(z)}$ and, for instance, $\frac{\delta}{\delta J^\mu(z)}$ respectively. In (19) $C_{\alpha\beta}$, $S_{\alpha\beta}$ and $T_{\alpha\beta}$ correspond respectively to the functional derivative representation of the canonical energy-momentum tensor $\hat{\Theta}^c_{\alpha\beta}$, the divergence $\partial^\lambda X_{\lambda\alpha\beta}$ of the covariant version of the tensor $X^{\lambda\alpha\beta}$ given in (11b) and, lastly, to a covariant version of the last term in (9a) but after necessary simplification. To verify the trace identity given in (10) it is enough to calculate $i\frac{\delta^3 Z}{\delta M_{\alpha\beta}(z)\delta J_\mu(s)\delta J_\nu(t)}\bigg|_{Sources=0}$ from (14) as that will yield, besides a multiplicative factor of $i^3$, the T*-product in the first term in eq.(10). On contraction with $g_{\alpha\beta}$ and simplification one immediately obtains the answer as

$$-i\delta(z-s)\langle 0|T^*\left(A^\mu(s)A^\nu(t)\right)|0\rangle_c - i\delta(z-t)\langle 0|T^*\left(A^\mu(s)A^\nu(t)\right)|0\rangle_c$$

thus verifying the trace identity in the unconstrained configuration. We have gone through the calculation above in a rather perfunctory fashion as there are no surprises to be expected here, but will exercise caution in the succeeding section as there will be qualitative changes in eqs.(17) and (18).

### 3. A check of eq.(10) with a boundary

A simple way to incorporate the Dirichlet boundary condition $A^\mu(t, x^1, x^2, x^3 = -a) = 0$ in to the generating functional given by (14) is to first rewrite it using the method of Bordag, Robaschik and Wieczorek[10] as

$$\exp iZ\left[J^\mu, K^{\mu\nu}, M^{\alpha\beta}\right] = N\int DA^\mu \delta\left(A^\lambda(t, x^1, x^2, x^3 = -a)\right)\exp i\int_x \left(L + J^\mu A_\mu + K^{\mu\nu}\partial_\mu A_\nu + M^{\alpha\beta}\hat{\Theta}_{\alpha\beta}\right)$$

(20)

which can be reworked as

$$\exp iZ\left[J^{\mu}, K^{\mu\nu}, M^{\alpha\beta}\right] = C\int DA^{\mu} Db_{\alpha}\, \exp i\int_{x}\left(L + J^{\mu}A_{\mu} + K^{\mu\nu}\partial_{\mu}A_{\nu} + M^{\alpha\beta}\hat{\Theta}_{\alpha\beta} + \delta\left(x^{3}+a\right)b_{\mu}A^{\mu}\right) \quad (21)$$

with C a normalizing factor and $b_{\lambda}(x)$ an auxiliary field that exists on the plate at $x^{3} = -a$ only; thus $b_{\lambda}(x)$ is a function of the three variables t, $x^{1}$ and $x^{2}$ only. One now rewrites (21) as

$$\exp iZ\left[J^{\mu}, K^{\mu\nu}, M^{\alpha\beta}\right] = \left(\sum_{0}^{\infty}\frac{i^{n}}{n!}(M^{\alpha\beta}\hat{\Theta}_{\alpha\beta})^{n}\right)N\int DA^{\mu} Db_{\alpha}\, \exp i\int_{x}\left(L + J^{\mu}A_{\mu} + K^{\mu\nu}\partial_{\mu}A_{\nu} + \delta\left(x^{3}+a\right)b_{\mu}A^{\mu}\right)$$

(22)

with L denoting the same Lagrangian as given by (8) but with $\lambda = -5$, and $\hat{\Theta}_{\mu\nu}$ now expressed in terms of functional derivatives as in (19). It is now easy to obtain from (22)

$$\exp iZ\left[J^{\mu}, K^{\mu\nu}, M^{\alpha\beta}\right] = \left(\sum_{0}^{\infty}\frac{i^{n}}{n!}(M^{\alpha\beta}\hat{\Theta}_{\alpha\beta})^{n}\right)N\int Db_{\alpha}\, \exp -\frac{i}{2}\int_{x}\int_{y}M_{\alpha}(x)D^{\alpha\beta}(x-y)M_{\beta}(y) \quad (23)$$

with
$$M_{\beta}(y) = J_{\beta}(y) + \overleftarrow{\partial}^{\mu}K_{\mu\beta}(y) + \delta\left(y^{3}+a\right)e^{-ik^{3}y^{3}}b_{\alpha}(y) \quad (24a)$$

and
$$M_{\alpha}(x) = J_{\alpha}(x) + K_{\mu\alpha}(x)\overrightarrow{\partial}^{\mu} + \delta\left(x^{3}+a\right)e^{ik^{3}x^{3}}b_{\alpha}(x) \quad (24b)$$

In eq.(23), $D^{\alpha\beta}(x-y)$ has the same form as in (16).

Note that the derivative operators in (24a) and (24b) are with respect to the variables $y^{\mu}$ and $x^{\mu}$ respectively. For the sake of clarity we rewrite the exponent in (23) below as:

$$\int_{xy}Q_{\alpha}(x)D^{\alpha\beta}(x-y)Q_{\beta}(y) + \int_{xy^{+}}Q_{\alpha}(x)D^{\alpha\beta}(x-y)b_{\beta}(y) + \int_{x^{+}y}b_{\alpha}(x)D^{\alpha\beta}(x-y)Q_{\beta}(y)$$
$$+ \int_{x^{+}y^{+}}b_{\alpha}(x)D^{\alpha\beta}(x-y)b_{\beta}(y) \quad (25)$$

with $Q_\alpha(x) \equiv J_\alpha(x) + K_{\beta\alpha}(x)\vec{\partial}^\beta$; also $\int_{xy^+}$ is a label for $\int d^4x \int d^4y\, \delta(y^3 + a)$, so that $\int_{x^+y^+}$ is shorthand for $\int d^4x\, \delta(x^3 + a) \int d^4y\, \delta(y^3 + a)$ and $\int_{xy}$ represents the familiar double integral without the Dirac $\delta$-function, with $y^3$ denoting the spatial z-component of the four vector $y^\mu$. With the definition

$$c_\alpha(x) \equiv b_\alpha(x) + \int_{zu^+} Q_\mu(z) D^{\mu\rho}(z-u) D^{-1}_{\rho\alpha}(u-x) \tag{26}$$

(25) becomes

$$\int\int_{x\,y} Q_\alpha(x)\left( D^{\alpha\mu}(x-y) - \int_{z^+u^+} D^{\alpha\beta}(x-z) D^{-1}_{\beta\rho}(z-u) D^{\rho\mu}(u-y) \right) Q_\mu(y) + \int\int_{x^+\,y^+} c_\alpha(x) D^{\alpha\beta}(x-y) c_\beta(y) \tag{27}$$

Eq.(32) below defines the inverse of $D^{\alpha\beta}(x-y)$ and (27) leads to

$$\exp iZ\left[J^\mu, K^{\mu\nu}, M^{\alpha\beta}\right] = \left( \sum_0^\infty \frac{i^n}{n!} (M^{\alpha\beta}\hat{\Theta}_{\alpha\beta})^n \right) \exp -\frac{i}{2} \int\int_{x\,y} Q_\alpha(x) \tilde{D}^{\alpha\beta}(x-y) Q_\beta(y) \tag{28}$$

with $\tilde{D}^{\alpha\beta}(x-y)$ denoting the expression in parentheses in (27). The second term in (27) is independent of the external sources and the resulting path integral due to the shift in (26) will be absorbed in to the normalization constant N. It is not difficult to check that in the presence of the boundary adopted in this paper, we obtain from (28)

$$i\frac{\delta^2 Z}{\delta J^\alpha(x)\delta J^\beta(y)}\bigg|_{sources=0} = i^2 \langle 0|T^*\left(A^\alpha(x) A^\beta(y)\right)|0\rangle_{c1} = -i\tilde{D}^{\alpha\beta}(x-y) \tag{29}$$

Eq.(29) thus defines the propagator for our calculation below; note that the connected Green's function has been given an additional subscript to distinguish it from that given in eq.(17). Further, it is the counterpart of eq.(2.17) in Bordag et al. [10] or to use a more recent reference, eq.(26) in Bordag and Lindig [11]. Using (18) one obtains with

$$\tilde{D}^{\alpha\beta}(x-y) = D^{\alpha\beta}(x-y) - \int_{z^+u^+} D^{\alpha\nu}(x-z)D^{-1}_{\nu\rho}(z-u)D^{\rho\beta}(u-y) \qquad (30)$$

$$\left(\partial^\mu\partial_\mu g_{\lambda\,\alpha} - \frac{6}{5}\partial_\lambda\partial_\alpha\right)\tilde{D}^{\alpha\beta}(x-y) = \delta^\beta_\lambda\delta^{(4)}(x-y) - \int_{z^+u^+}\delta^\nu_\lambda\delta(x-z)D^{-1}_{\nu\rho}(z-u)D^{\rho\beta}(u-y)$$

$$= \delta^\beta_\lambda\delta^{(4)}(x-y) - \int_{u^+}\delta\left(x^3+a\right)D^{-1}_{\lambda\rho}(x-u)D^{\rho\beta}(u-y)$$

$$= \delta^\beta_\lambda\delta^{(4)}(x-y) - \delta^\beta_\lambda\delta^{(3)}(x-y)\delta\left(x^3+a\right) \qquad (31a)$$

when $y^3 = -a$; or, to rewrite (31a)

$$\left(\partial^\mu\partial_\mu g_{\lambda\,\alpha} - \frac{6}{5}\partial_\lambda\partial_\alpha\right)\tilde{D}^{\alpha\beta}(x-y) = \delta^\beta_\lambda\delta^{(4)}(x-y) - \delta^\beta_\lambda\delta^{(3)}(x-y)\delta\left(x^3-y^3\right) \qquad (31b)$$

when $y^3 = -a$. While the first term in (31b) arises from the r.h.s. of eq (18), the latter being associated with $D^{\alpha\beta}(x-y)$, the second term despite its resemblance to the first is valid only when $y^3 = -a$, i.e. on the single plate introduced in this paper. Indeed, it is the latter term that will play a decisive role in invalidating the trace identity as given by (10). The inverse of $D^{\alpha\beta}(x-y)$, namely, $D^{-1}_{\rho\alpha}(x-y)$ satisfies the relation

$$\int_{y^+} D^{-1}_{\rho\alpha}(u-y)D^{\alpha\beta}(y-z) = \delta^\beta_\rho\delta^{(3)}(u-z) \qquad (32)$$

when $u^3 = -a$ and $z^3 = -a$; and the form of $D^{-1}_{\rho\alpha}(x-y)$ is worked out in the Appendix to this paper. Shelving these details it is enough to anticipate here that given the following form for $D^{\alpha\beta}(x-y)$ viz.,

$$D^{\alpha\beta}(x-y) = i\int_k e^{-ik\cdot(x-y)-i|\vec{k}|\left|(x^3-y^3)\right|}\frac{1}{|\vec{k}|}\left(\frac{1}{2}\eta^{\alpha\beta} - \rho\frac{q^\alpha q^\beta}{|\vec{k}|^2}\right) \qquad (33)$$

with: i. the momentum integration being 3-dimensional so that $\int_k$ denotes $\int\frac{d^3k}{(2\pi)^3}$

ii. $\eta^{\mu\nu}$ being a diagonal matrix given by $\eta^{\mu\nu} = (1,-1,-1,5)$

iii. $q^{\mu} = (-k^0, k^1, k^2, -|\vec{k}|)$ for $\mu = 0,1,2$ and 3, and,

iv. $\rho = \begin{cases} \lambda_0, & \alpha \neq 3, \beta \neq 3 \\ \lambda_0 - \dfrac{3}{2}, & \alpha \neq 3, \beta = 3 \end{cases}$, $\lambda_0 \equiv \dfrac{3}{2}\left(1 + |\vec{k}||x^3 - y^3|\right)$

one can show that

$$D_{\alpha\beta}^{-1}(x-y) = -i\int_p e^{-ik\cdot(x-y)+i|\vec{k}||x^3-y^3|} |\vec{k}| \left(b_{\alpha\beta} + c\frac{p_\alpha p_\beta}{|\vec{k}|^2}\right) \tag{34}$$

with:

i. $b_{\alpha\beta}$ being a diagonal matrix given by $b_{\alpha\beta} = (2,-2,-2)$ for $\alpha, \beta = 0, 1$ and 2

ii. $p_\alpha = \left(5\lambda_0 k^0, 5\lambda_0 k^1, 5\lambda_0 k^2, \left(\lambda_0 - \dfrac{3}{2}\right)|\vec{k}|\right)$ and, iii. $c = \dfrac{1}{15\lambda_0\left(\dfrac{2}{3} - \lambda_0\right)}$

with $\lambda_0$ defined as above. Note that eqs.(33) and (34) define $D^{\alpha\beta}(x-y)$ and $D_{\alpha\beta}^{-1}(x-y)$ on the three dimensional subspace $x^\mu$ ($\mu = 0, 1, 2$) only. Also eqs. (33) and (34) are the respective counterparts of eqs. (2.10) and (2.14) of Ref. 10.

We shall now take up the verification of the trace identity in the presence of the boundary remembering – see eq.(16) – that $D^{\alpha\beta}(x-y) = D^{\beta\alpha}(x-y)$. The steps below use this symmetry property as well as eqs.(30) and (31b) ; they *do not use* eqs.(33) and (34) for $D^{\alpha\beta}(x-y)$ and $D_{\alpha\beta}^{-1}(x-y)$ respectively. In other words, the main conclusions of this paper do not depend on eqs.(33) and (34), with the latter as well as the content of the Appendix to this paper only serving the limited purpose of working out a possible form for $D_{\alpha\beta}^{-1}(x-y)$, at least partly inspired by the work of Bordag et al.[10] and Bordag and Lindig[11] and as a complement to our work in Ref.9.

As in the previous section it is enough to calculate $i\dfrac{\delta^3 Z}{\delta M_{\alpha\beta}(z)\delta J_\mu(s)\delta J_\nu(t)}\bigg|_{Sources=0}$ from (28) as that will yield, besides a multiplicative factor of $i^3$, the T* - product in the first term in eq.(10). To make the discussion reader-friendly, we shall now present below the results got from each of the three terms in (19), but after contraction with $g_{\alpha\beta}$, remembering that the partial derivative operators are with respect to the variable $z^\mu$:

Term 1:

$$2i\left\{\partial^\alpha \tilde{D}^\rho_\mu(z-s)\partial_\alpha \tilde{D}_{\rho\nu}(z-t) - \partial^\alpha \tilde{D}^\rho_\mu(z-s)\partial_\rho \tilde{D}_{\alpha\nu}(z-t) - \frac{1}{5}\partial^\alpha \tilde{D}_{\alpha\mu}(z-s)\partial^\rho \tilde{D}_{\rho\nu}(z-t)\right\}$$

Term 2:

$$2i\left\{\begin{array}{l}\dfrac{6}{5}\partial^\lambda \tilde{D}_{\lambda\mu}(z-s)\partial^\beta \tilde{D}_{\beta\nu}(z-t) + \partial^\lambda \tilde{D}_{\beta\mu}(z-s)\left(\partial^\beta \tilde{D}_{\lambda\nu}(z-t) - \partial_\lambda \tilde{D}^\beta_{\;\nu}(z-t)\right)\\[2mm] +\tilde{D}_{\lambda\mu}(z-s)\left(\dfrac{8}{5}\partial^\lambda\partial^\beta \tilde{D}_{\beta\nu}(z-t) - \partial^\beta\partial_\beta \tilde{D}^\lambda_{\;\nu}(z-t)\right) + \tilde{D}_{\lambda\nu}(z-t)\left(\dfrac{8}{5}\partial^\lambda\partial^\beta \tilde{D}_{\beta\mu}(z-s) - \partial^\beta\partial_\beta \tilde{D}^\lambda_{\;\mu}(z-s)\right)\\[2mm] +\partial^\lambda \tilde{D}_{\alpha\nu}(z-t)\left(\partial^\alpha \tilde{D}_{\lambda\mu}(z-s) - \partial_\lambda \tilde{D}^\alpha_{\;\mu}(z-s)\right)\end{array}\right\}$$

Adding the two terms above results in

$$2i\left\{\begin{array}{l}\partial^\alpha \tilde{D}_{\alpha\mu}(z-s)\partial^\beta \tilde{D}_{\beta\nu}(z-t) + \partial^\lambda \tilde{D}_{\alpha\nu}(z-t)\left(\partial^\alpha \tilde{D}_{\lambda\mu}(z-s) - \partial_\lambda \tilde{D}^\alpha_{\;\mu}(z-s)\right)\\[2mm] \tilde{D}_{\lambda\mu}(z-s)\left(\dfrac{8}{5}\partial^\lambda\partial^\beta \tilde{D}_{\beta\nu}(z-t) - \partial^\beta\partial_\beta \tilde{D}^\lambda_{\;\nu}(z-t)\right) + \tilde{D}_{\lambda\nu}(z-t)\left(\dfrac{8}{5}\partial^\lambda\partial^\beta \tilde{D}_{\beta\mu}(z-s) - \partial^\beta\partial_\beta \tilde{D}^\lambda_{\;\mu}(z-s)\right)\end{array}\right\}$$

Finally, we have for

Term 3: $i\left\{\partial^\alpha\partial_\alpha\left(\tilde{D}_{\rho\mu}(z-s)\tilde{D}^\rho_{\;\nu}(z-t)\right) - 2\partial^\alpha\partial^\beta\left(\tilde{D}_{\alpha\mu}(z-s)\tilde{D}_{\beta\nu}(z-t)\right)\right\}$

The addition of Term 3 to the sum preceding it now yields the final answer for the first term in the trace identity given by (10), viz.

$$i\left\{\tilde{D}_{\lambda\mu}(z-s)\left(\frac{6}{5}\partial^{\lambda}\partial^{\beta}\tilde{D}_{\beta\nu}(z-t)-\partial^{\beta}\partial_{\beta}\tilde{D}^{\lambda}{}_{\nu}(z-t)\right)+\tilde{D}_{\lambda\nu}(z-t)\left(\frac{6}{5}\partial^{\lambda}\partial^{\beta}\tilde{D}_{\beta\mu}(z-s)-\partial^{\beta}\partial_{\beta}\tilde{D}^{\lambda}{}_{\mu}(z-s)\right)\right\}$$

(35)

Note that the first entry in the parentheses in (35) becomes

$$\frac{6}{5}\partial^{\lambda}\partial^{\beta}\tilde{D}_{\beta\nu}(z-t)-\partial^{\beta}\partial_{\beta}\tilde{D}^{\lambda}{}_{\nu}(z-t) = -\left(\partial^{\beta}\partial_{\beta}\tilde{D}^{\lambda}{}_{\nu}(z-t)-\frac{6}{5}\partial^{\lambda}\partial^{\beta}\tilde{D}_{\beta\nu}(z-t)\right)$$

$$= -\left(\partial^{\beta}\partial_{\beta}g^{\lambda\rho}-\frac{6}{5}\partial^{\lambda}\partial^{\rho}\right)\tilde{D}_{\rho\nu}(z-t)$$

$$= -\left(\delta^{\lambda}_{\nu}\delta^{(4)}(z-t)-\delta^{\lambda}_{\nu}\delta^{(3)}(z-t)\delta(z^{3}-t^{3})\right) \quad (36a)$$

with the second entry similarly given by

$$\frac{6}{5}\partial^{\lambda}\partial^{\beta}\tilde{D}_{\beta\mu}(z-s)-\partial^{\beta}\partial_{\beta}\tilde{D}^{\lambda}{}_{\mu}(z-s) = -\left(\delta^{\lambda}_{\mu}\delta^{(4)}(z-s)-\delta^{\lambda}_{\mu}\delta^{(3)}(z-s)\delta(z^{3}-s^{3})\right) \quad (36b)$$

With eqs.(36a) and (36b), (35) now becomes the sum of

$$R \equiv -i\tilde{D}_{\nu\mu}(t-s)\delta^{(4)}(z-t)-i\tilde{D}_{\mu\nu}(s-t)\delta^{(4)}(z-s)$$

$$= -i\tilde{D}_{\nu\mu}(t-s)\left(\delta^{(4)}(z-t)+\delta^{(4)}(z-s)\right) \quad (37)$$

and

$$Q \equiv i\tilde{D}_{\nu\mu}(t-s)\delta^{(3)}(z-t)\delta(z^{3}-t^{3}) + i\tilde{D}_{\mu\nu}(s-t)\delta^{(3)}(z-s)\delta(z^{3}-s^{3})$$

$$= i\tilde{D}_{\nu\mu}(t-s)\left(\delta^{(3)}(z-t)\delta(z^{3}-t^{3})+\delta^{(3)}(z-s)\delta(z^{3}-s^{3})\right) \quad (38)$$

Thus the trace identity given by (10) now becomes

$$g_{\alpha\beta}\left\langle 0\left|T^{*}\left(\hat{\Theta}^{\alpha\beta}(z)A^{\mu}(s)A^{\nu}(t)\right)\right|0\right\rangle_{c} = i(Q+R)$$

i.e.,

$$g_{\alpha\beta}\langle 0|T^*\left(\hat{\Theta}^{\alpha\beta}(z)A^\mu(s)A^\nu(t)\right)|0\rangle_c + \langle 0|T^*\left(A^\alpha(x)A^\beta(y)\right)|0\rangle_{c1}\left(i\delta^{(4)}(z-t)+i\delta^{(4)}(z-s)\right)$$

$$= iQ \tag{39}$$

with $\quad iQ = \langle 0|T^*\left(A^\mu(s)A^\nu(t)\right)|0\rangle_{c1}\left(i\delta^{(3)}(z-t)\delta(z^3-t^3)+i\delta^{(3)}(z-s)\delta(z^3-s^3)\right)$ (39a)

Clearly, the r.h.s. of (39) is *not zero* on the plate at $z^3 = -a$ and the trace identity given by (10) is thus invalid on the single plate we have introduced as a boundary in this paper. Note that (39a) is formally similar to and with the same sign as the second term on the l.h.s. of the same equation but originates from the second term on the r.h.s. of eq.(30). We shall now view the above result in another context below, namely, as a support to our work in Ref.9

**Discussion:**

In our earlier report [9] associated with the Lagrangian density of a real massive noninteracting scalar field in 2 + 1 dimensions, viz.

$$L = \frac{1}{2}\partial^\mu\phi\,\partial_\mu\phi - \frac{1}{2}m^2\phi^2 \tag{40}$$

we had dealt with the trace identity

$$g_{\alpha\beta}\langle 0|T^*\left(\hat{\Theta}^{\alpha\beta}(z)\phi(s)\phi(u)\right)|0\rangle_c + \frac{1}{2}i\delta^{(3)}(z-s)\langle 0|T^*\left(\phi(s)\phi(u)\right)|0\rangle_c$$
$$+ \frac{1}{2}i\delta^{(3)}(z-u)\langle 0|T^*\left(\phi(s)\phi(u)\right)|0\rangle_c = m^2\langle 0|T^*\left(\phi^2(z)\phi(s)\phi(u)\right)|0\rangle_c \tag{41}$$

To make our point, let's recall here from Ref.9 that while eq.(41) held in the unconstrained configuration, it did not on a boundary on which the Dirichlet boundary condition

$\phi(t,x^1,x^2=-a)=0$ was imposed. In other words, in the latter case (41) was modified to

$$g_{\alpha\beta}\langle 0|T^*\left(\hat{\Theta}^{\alpha\beta}(z)\phi(s)\phi(u)\right)|0\rangle_c + \frac{1}{2}i\delta^{(3)}(z-s)\langle 0|T^*\left(\phi(s)\phi(u)\right)|0\rangle_{c1}$$
$$+\frac{1}{2}i\delta^{(3)}(z-u)\langle 0|T^*\left(\phi(s)\phi(u)\right)|0\rangle_{c1} = m^2\langle 0|T^*\left(\phi^2(z)\phi(s)\phi(u)\right)|0\rangle_c + iK \quad (42)$$

where

$$iK = -\frac{1}{2}\langle 0|T^*\left(\phi(s)\phi(u)\right)|0\rangle_{c1}\left\{-i\delta^{(2)}(z-u)\delta(z^2-u^2)-i\delta^{(2)}(z-s)\delta(z^2-s^2)\right\} \quad (43)$$

with $z^2$ being the spatial y-component of the 3-vector $z^\mu$, with the same understanding for $u^2$ and $s^2$.

Eq.(42) is thus the counterpart of the r.h.s. of (39), but for the Lagrangian in 2 + 1 dimensions given by (40). Apart from a difference of the canonical scale dimension which is respectively $\frac{1}{2}$ for the scalar field and appears in (42) and (43), and 1 for the vector field as seen in (39) and (39a), both eqs.(39a) and (43) are similar despite the fact that the two equations refer in details to two different theories.

Put differently, the observed violation in Ref.9 of the naïve trace identity given by (41) above on a plate on which the Dirichlet boundary condition has been imposed is now sustained by the failure of the counterpart of eq.(41) - namely eq.(10) – on a plate on which a similar boundary condition is imposed in this paper. Both these naïve identities are otherwise, needless to say so, maintained in free space. For the sake of completeness, we give below the form of the canonical dilatation current for the Lagrangian given by (40), viz.

$$D_c^\lambda = x_\mu \hat{\Theta}_c^{\lambda\mu} + \frac{1}{4}\partial^\lambda \phi^2 \quad (44)$$

while the counterpart of (9b) is

$$X^{\alpha\beta\mu\nu} = \frac{1}{4}\left(g^{\alpha\beta}g^{\mu\nu} - g^{\alpha\nu}g^{\beta\mu}\right)\phi^2 \quad (44a)$$

Eqs.(44) and (44a) together lead as shown in Ref.9 to a modified version of the dilatation current namely, $D^\mu = x_\nu \hat{\Theta}^{\mu\nu}$ with $\hat{\Theta}_{\mu\nu}$ defined in terms of $\hat{\Theta}_c^{\mu\nu}$ by eq. (9a) but with $X^{\alpha\beta\mu\nu}$ as given by (44a).

To conclude, the twin checks on the validity of the trace identity given by (10) in Secs. 2 and 3 of this paper for the Lagrangian of the noninteracting electromagnetic field given by (8) in the covariant gauge, reinforce our conclusions obtained in the earlier paper[9] that the trace identity associated with the scale transformation $x^\mu \to x'^\mu = e^{-\rho} x^\mu$ for the Lagrangian (40) becomes anomalous – in the sense that it is not maintained – when one moves from an unconstrained configuration where the identity is respected to a constrained one, the latter being characterised by a boundary on which a Dirichlet boundary condition is imposed; a notable feature of this anomalous term being the appearance of the canonical scale dimension as a numerical coefficient . As another example of the latter observation, let us reconsider the Lagrangian given by (40) in 1 + 1 dimensions in which case the canonical scale dimension of the scalar field becomes zero, and the trace identity given by (41) now reads as

$$g_{\alpha\beta} \langle 0 | T^* \left( \hat{\Theta}^{\alpha\beta}(z) \phi(s) \phi(u) \right) | 0 \rangle_c = m^2 \langle 0 | T^* \left( \phi^2(z) \phi(s) \phi(u) \right) | 0 \rangle_c \qquad (45)$$

with the canonical dilatation current now given by $D_c^\alpha = x_\mu \hat{\Theta}_c^{\alpha\mu}$; the absence of the field virial in this case, in the sense that it is zero, now implies that: One, $\hat{\Theta}^{\mu\nu} = \hat{\Theta}_c^{\mu\nu}$ with $\hat{\Theta}_c^{\mu\nu}$ being the canonical energy-momentum tensor, and secondly, while the trace identity given by (44) is respected in the unconstrained configuration, it is also maintained in the case when one introduces a plate on which the Dirichlet condition $\phi(t, x^1 = -a) = 0$ holds. We refer the interested reader to Ref.9 for a somewhat more detailed discussion on these issues and conclude this paper below with an Appendix wherein a fairly detailed evaluation of $D_{\alpha\beta}^{-1}(x-y)$ is presented.

## Appendix

We begin with the eq.(16) of the text, namely,

$$iD^{\alpha\beta}(x-y) = i\int_k \frac{e^{-ik\cdot(x-y)}}{k^2+i\varepsilon}\left(6\frac{k^\alpha k^\beta}{k^2}-g^{\alpha\beta}\right) \tag{A1}$$

and perform a Wick rotation to obtain for

$$D^{00}(x-y) = -i\int_k e^{-ik\cdot(x-y)}\left(6(k^0)^2\int_0^\infty re^{-rk^2}dr - g^{00}\int_0^\infty e^{-rk^2}dr\right) \tag{A2}$$

We shall now integrate over $k^3$ in (A2) and get after doing the r – integral the result

$$D^{00}(x-y) = -i\int_k e^{-ik\cdot(x-y)-|\vec{k}||x^3-y^3|}\left(-g^{00}\frac{1}{2|\vec{k}|}+\frac{3(k^0)^2}{2|\vec{k}|^3}\left(1+|\vec{k}||x^3-y^3|\right)\right)$$

A similar effort on the other components of $D^{\alpha\beta}(x-y)$ leads to

$$D^{0j}(x-y) = (i)^2\int_k e^{-ik\cdot(x-y)-|\vec{k}||x^3-y^3|}\frac{3k^0 k^j}{2|\vec{k}|^3}\left(1+|\vec{k}||x^3-y^3|\right) \tag{A3}$$

for $j \neq 3$. Likewise for $D^{ij}(x-y)$ for $i \neq 3$, $j \neq 3$ we get

$$D^{ij}(x-y) = i\int_k e^{-ik\cdot(x-y)-|\vec{k}||x^3-y^3|}\left(g^{ij}\frac{1}{2|\vec{k}|}+\frac{3k^i k^j}{2|\vec{k}|^3}\left(1+|\vec{k}||x^3-y^3|\right)\right) \tag{A4}$$

It is not difficult now to obtain a general expression for $D^{\alpha\beta}(x-y)$ after undoing the Wick rotation and this is given by

$$D^{\alpha\beta}(x-y) = i\int_k e^{-ik\cdot(x-y)-i|\vec{k}||(x^3-y^3)|}\frac{1}{|\vec{k}|}\left(\frac{1}{2}\eta^{\alpha\beta}-\rho\frac{q^\alpha q^\beta}{|\vec{k}|^2}\right) \tag{A5}$$

with : the momentum integration being 3-dimensional so that $\int_k$ denotes $\int \frac{d^3k}{(2\pi)^3}$, $\eta^{\mu\nu}$ being a diagonal matrix given by $\eta^{\mu\nu} = (1,-1,-1,5)$, $q^\mu = (-k^0, k^1, k^2, -|\vec{k}|)$ for $\mu = 0,1,2$ and 3, and, $\rho = \begin{cases} \lambda_0, & \alpha \neq 3, \beta \neq 3 \\ \lambda_0 - \frac{3}{2}, & \alpha \neq 3, \beta = 3 \end{cases}$ where $\lambda_0 \equiv \frac{3}{2}\left(1 + |\vec{k}||x^3 - y^3|\right)$.

We now use the following form for $D_{\alpha\beta}^{-1}(x-y)$

$$D_{\alpha\beta}^{-1}(x-y) = -i\int_p e^{-ip\cdot(x-y) + i|\vec{p}||(x^3-y^3)|} |\vec{p}| \left(d_{\alpha\beta} + b\frac{r_\alpha r_\beta}{|\vec{p}|^2}\right) \tag{A6}$$

so as to reflect the symmetry properties of $D^{\alpha\beta}(x-y)$ exhibited by eq.(16) and mentioned earlier in the text of the paper. The $d_{\alpha\beta}$, b and $r_\alpha$ will be fixed by using the definition

$$\int_{z^+} D_{\alpha\beta}^{-1}(x-z) D^{\beta\rho}(z-y) = \delta_\alpha^\rho \delta^{(3)}(x-y) \tag{A7}$$

when $x^3 = -a$ and $y^3 = -a$; $z^+$ being a symbol for $\int d^4z\, \delta(z^3 + a)$ and $z^3$ being the spatial -z component of the 4-vector $z^\mu$ with a similar understanding for $x^\mu$ and $y^\mu$, it being additionally understood that $d_{\alpha\beta}$ is a diagonal matrix. Repeated use of (A7) for $\alpha = 1, \rho = 0$, 2 and 3 enable us to arrive at $r_0 = 5\lambda_0 k^0$, $r_2 = 5\lambda_0 k^2$ and $r_3 = \left(\lambda_0 - \frac{3}{2}\right)|\vec{k}|$. A similar exercise with $\alpha = 0$ and 2 but with $\rho = 0, 1, 2$ and 3 yields $d_{00} = 2$, $d_{11} = -2 = d_{22}$, $r_1 = 5\lambda_0 k^1$ and $b = \dfrac{1}{15\lambda_0\left(\dfrac{2}{3} - \lambda_0\right)}$.

# References

1. S. Coleman and R. Jackiw, Ann.Phys.(N.Y.) **67**, 552(1971).

2. See remarks following eq.(A.16) in Ref.1

3. See eq.(A.18) in Ref.1

4. H. B.G. Casimir, Proc. Kon. Ned. Akad. Wetensch., **51**,793(1948);for additional references see K. A. Milton, The Casimir effect :Physical Manifestations of Zero-Point Energy, World Scientific(New Jersey,2001).

5. G. Plunien, B. Muller and W. Greiner, Phys. Rep.**134**,87(1986).

6. M.J. Spaarnay, Physica **24**,751(1958).

7. See also, M. Bordag, U. Mohideen and V. M. Mostepanenko,

Phys.Rep.**353**,1(2001) for additional references.

8. H. B.G. Casimir,Ref.4

9. S.G. Kamath, submitted for publication.

10. M. Bordag, D. Robaschik and E. Wieczorek, Ann. Phys.(N.Y.)**165**,192(1985).

11. M. Bordag and J. Lindig, Phys. Rev. D **58**,045003(1998).